# Improving the efficiency of quantum engineering of SCSs by adding two demultiplexed input photons

Mikhail S. Podoshvedov[1,2] and Sergey A. Podoshvedov[1,2*]

[1]*Laboratory of quantum information processing and quantum computing, laboratory of quantum engineering of light, South Ural State University (SUSU), Lenin Av. 76, Chelyabinsk, Russia*
[2]*Laboratory of quantum engineering of light, South Ural State University (SUSU), Lenin Av. 76, Chelyabinsk, Russia*
[*]<u>sapodo68@gmail.com</u>

**Abstract:** Conditional addition and subtraction of photons is a powerful tool for quantum engineering of continuous variable (CV) states that forms the fundamental building blocks of advanced photonic technologies. For high fidelity information processing, precise control of quantum engineering of CV states is highly demanded. We propose a scheme for measurement induced quantum engineering of even/odd superposition of coherent states (SCSs) of amplitude$> 2.5$ and with fidelity exceeding$> 0.99$. It includes single-mode squeezed vacuum (SMSV) state and two separate photons. The output parameters of the SCSs are tuned using initial squeezing of the SMSV states and the beam splitter (BS) parameter and depend on number of subtracted photons in two measurement channels. The introduction of two demultiplexed photons is a key element to improving the efficiency of quantum engineering of SCSs. When using two additional photons, both an increase in the fidelity of the output CV state and a gain in probability at least by an order of magnitude compared to the case without auxiliary input photons are observed.

## 1. Introduction

Controlled quantum engineering of non-classical states of light is key to the rapid development of new optical quantum technologies. Standard optical transformations are based on Gaussian operations such as squeezing, BS mixing and displacement [1,2]. Gaussian states, such as vacuum and thermal states, occur in nature or, like coherent and squeezed states, can be produced experimentally on demand. Despite the sufficient availability of the Gaussian states, the range of their practical applications may be limited. So, achieving fault-tolerant quantum computing is not possible using Gaussian states and operations alone [3].

The range of applications of non-Gaussian states and non-Gaussian operations is wider, but nevertheless, generating non-Gaussian states with high fidelity and large probability remains experimentally challenging task [4]. If, after mixing the Gaussian states of light, a portion of the light energy is directed into the auxiliary measurement mode, the photon number measurement provides effective nonlinearity and serves as a key tool for preparing highly nonclassical optical quantum states beyond the realm of Gaussian operations [5-8]. In addition to the photon subtraction, photon addition can also be used [9-12]. Conditional addition and subtraction of photons are powerful and useful tools that find wide application in quantum photonics. They involve quantum engineering of optical nonclassical states [13-25] including generation of single mode superposition of coherent states (SCSs) [26-31]. Using the combination of photon addition and subtraction, an experimental test of quantum commutation relation was demonstrated in [32]. Possible application of photon addition and subtraction technologies also includes preparation of hybrid entangled states [33-36].

For advanced applications, state engineering with adjustable parameters by means of simultaneous addition and subtraction of photons is indispensable. Squeezed light [37,38] is an important nonclassical resource and the starting ingredient on which the engineering of new CV states relies. Photon subtraction from SMSV state is a standard method in SCSs generation but the increase in its output amplitude is limited by the rate of its generation, which decreases significantly with an increase in the number of subtracted photons [29-31]. Even with the use of squeezing and beam splitter parameter optimization, the probability of the target CV state remains low [8]. Therefore, more efficient methods to implement large amplitude even/odd SCSs with high fidelity must be developed for their further practical application.

In this paper, we develop an approach for more efficient generation of SCSs by adding two input separate photons. The addition of two demultiplexed photons is achieved using a system of two identical beam splitters through which the SMSV state passes. Optimization by the BS parameter and initial squeezing allows us to find the maximum fidelity of the target CV state. In addition to the improved fidelity of the output CV state, it turns out that the probability of generating the target state increases by an order of magnitude compared to the case without input photons. Taking into account the possibilities of implementing the SMSV states [39,40] and the progress in photon number resolving detectors based on TES technology [41], the proposed method is feasible in practice.

## 2. Quantum engineering of SCSs states by demultiplexing input photons

We consider the conditional CV state realized by passing the SMSV state

$$|SMSV(y)\rangle = \frac{1}{\sqrt{\cosh s}}\sum_{n=0}^{\infty} \frac{y^n}{\sqrt{(2n)!}} \frac{(2n)!}{n!} |2n\rangle \quad (1)$$

through a system of two identical beam splitters located one after the other in a row as shown in Figure 1, followed by measuring the number of photons by two PNR detectors. Here, squeezing parameter $y = \tanh s/2$ is used determined through the squeezing amplitude $s > 0$. Since the squeezing amplitude can in principle take values $0 \leq s < \infty$, the squeezing parameter is in the range $0 \leq y < 0.5$. Another way to characterize the SMSV state is related to the squeezing expressed in decibels $S = -10 \log(\exp(-2s))\, dB$, as well as the average number of photons $\langle n_{SMSV} \rangle = \sinh^2 s$.

As can be seen from Figure 1, SMSV states occupies the first mode and the auxiliary single photons in the additional modes 2 and 3 of the system are mixed with the original CV state. Mode 1 is common to both beam splitters. The beam splitters in Figure 1 are chosen to ensure that the input creation operators $\hat{a}_1^\dagger$, $\hat{a}_2^\dagger$ and $\hat{a}_3^\dagger$ are transformed as $\hat{U}_{12}\hat{a}_1^\dagger \hat{U}_{12}^\dagger = t\hat{a}_1^\dagger - r\hat{a}_2^\dagger$, $\hat{U}_{12}\hat{a}_2^\dagger \hat{U}_{12}^\dagger = r\hat{a}_1^\dagger + t\hat{a}_2^\dagger$, $\hat{U}_{13}\hat{a}_1^\dagger \hat{U}_{13}^\dagger = t\hat{a}_1^\dagger - r\hat{a}_3^\dagger$ and $\hat{U}_{13}\hat{a}_3^\dagger \hat{U}_{13}^\dagger = r\hat{a}_1^\dagger + t\hat{a}_3^\dagger$, where $\hat{U}_{1i}$ with $i = 1,2$ is the unitary operator of the beam splitter and $\hat{U}_{12}^\dagger$ is the operator Hermitian adjoint to the original $\hat{U}_{1i}$ with arbitrary real transmittance $t > 0$ and reflectance $r > 0$ subject to the normalization condition $t^2 + r^2 = 1$ for both BSs. The set of the BSs transforms the original SMSV state in Eq. (1) into a hybrid entangled state [29,31]

$$BS_{13}BS_{12}(|SMSV\rangle_1 |1\rangle_2 |1\rangle_3) = \frac{1}{\sqrt{\cosh s}}$$

$$\begin{pmatrix} c_0^{(1)}(y_1,B)c_0^{(1)}(y_2,B)\sqrt{G_{00}^{(11)}(y_2,B)}|\Psi_{00}^{(11)}(y_2,B)\rangle_1 |0\rangle_2 |0\rangle_3 + \\ c_0^{(1)}(y_1,B)\left(\sum_{k_2=0}^{\infty} c_{k_2}^{(1)}(y_2,B)\sqrt{G_{0k_2}^{(11)}(y_2,B)}|\Psi_{0k_2}^{(11)}(y_2,B)\rangle_1 |k_2\rangle_3\right)|0\rangle_2 + \\ c_0^{(1)}(y_2,B)\left(\sum_{k_1=0}^{\infty} c_{k_1}^{(1)}(y_1,B)\sqrt{G_{k_10}^{(11)}(y_2,B)}|\Psi_{k_10}^{(11)}(y_2,B)\rangle_1 |k_1\rangle_2\right)|0\rangle_3 + \\ \sum_{k_1=1}^{\infty}\sum_{k_2=1}^{\infty} c_{k_1}^{(1)}(y_1,B)c_{k_2}^{(1)}(y_2,B)\sqrt{G_{k_1k_2}^{(11)}(y_2,B)}|\Psi_{k_1k_2}^{(11)}(y_2,B)\rangle_1 |k_1\rangle_2 |k_2\rangle_3 \end{pmatrix}, \quad (2)$$

The conclusion of the state in equation (2) follows from the general methodological approach developed in [29].

Here, the amplitudes with superscript (1) responsible for single input photon are given by

$$c_k^{(1)}(y,B) = \frac{1}{\sqrt{1+B}}\begin{cases} \sqrt{B}, & \text{if } k = 0 \\ (-1)^{k+1}\frac{(yB)^{\frac{k-1}{2}}}{\sqrt{k!}}k, & \text{if } k \neq 0 \end{cases}, \qquad (3)$$

with a subscript $k$ corresponding to the number of subtracted photons. The product of the amplitudes of $c_k^{(1)}(y_1, B)$ and $c_k^{(1)}(y_2, B)$ depend on three parameters $y_1$, $y_2$ and $B$. As for the beam splitter parameter $B$, it is determined through the ratio transmission $T = t^2$ and reflection coefficients $R = r^2$ as $B = (1-t^2)/t^2$, which allows us to derive them through it as $T = 1/(1+B)$ and $R = B/(1+B)$, respectively. From the definition it follows that the value $B = 1$ corresponds to a balanced BS with $T = R = 0.5$, while $B < 1$ is responsible for transmitting beam splitter beam splitter with $T > R$ and the value $B > 1$ describes a more reflective beam splitter with $R > T$. A parameter $y_1$ is the initial squeezing parameter $y$ reduced by $t^2$ times, which the amplitude acquires after passing the first BS, that is, $y_1 = yt^2 = y/(1+B) \leq y$. After passing through the second BS, the squeezing parameter also decreases by $t^2$ times, that is, it becomes $y_2 = y_1 t^2 = y_1/(1+B) = yt^4 = y/(1+B)^2$.

As for the CV states of a certain parity, we present only those of them that are used in in subsequent numerical modeling, i.e. that is, for which the number of subtracted photons $k_1 > 0$ and $k_2 > 0$

$$|\Psi_{2m_1\,2m_2}^{(11)}(y_2,B)\rangle = \frac{1}{\sqrt{G_{2m_1\,2m_2}^{(11)}(y_2,B)}}\sum_{n=0}^{\infty}\begin{pmatrix}\frac{y_2^n}{\sqrt{(2n)!}}\frac{(2(n+m_1+m_2-1))!}{(n+m_1+m_2-1)!}\\\left(1-\frac{2m_2-1}{2m_1}B-\frac{B}{2m_1}2n\right)\\\left(1-\frac{B}{2m_2}2n\right)\end{pmatrix}|2n\rangle, \qquad (4)$$

$$|\Psi_{2m_1\,2m_2+1}^{(11)}(y_2,B)\rangle = \sqrt{\frac{y_2}{G_{2m_1\,2m_2+1}^{(11)}(y_2,B)}}\sum_{n=0}^{\infty}\begin{pmatrix}\frac{y_2^n}{\sqrt{(2n+1)!}}\frac{(2(n+m_1+m_2))!}{(n+m_1+m_2)!}\\\left(1-\frac{2m_2}{2m_1}B-\frac{B}{2m_1}(2n+1)\right)\\\left(1-\frac{B}{2m_2+1}(2n+1)\right)\end{pmatrix}|2n+1\rangle, (5)$$

$$|\Psi_{2m_1+1\,2m_2}^{(11)}(y_2,B)\rangle = \sqrt{\frac{y_2}{G_{2m_1+1\,2m_2}^{(11)}(y_2,B)}}\sum_{n=0}^{\infty}\begin{pmatrix}\frac{y_2^n}{\sqrt{(2n+1)!}}\frac{(2(n+m_1+m_2))!}{(n+m_1+m_2)!}\\\begin{pmatrix}1-\frac{2m_2-1}{2m_1+1}B-\\\frac{B}{2m_1+1}(2n+1)\end{pmatrix}\\\left(1-\frac{B}{2m_2}(2n+1)\right)\end{pmatrix}|2n+1\rangle, \qquad (6)$$

$$|\Psi_{2m_1+1\,2m_2+1}^{(11)}(y_2,B)\rangle = \frac{1}{\sqrt{G_{2m_1+1\,2m_2+1}^{(11)}(y_2,B)}}\sum_{n=0}^{\infty}\begin{pmatrix}\frac{y_2^n}{\sqrt{(2n)!}}\frac{(2(n+m_1+m_2))!}{(n+m_1+m_2)!}\\\begin{pmatrix}1-\frac{2m_2}{2m_1+1}B-\\\frac{B}{2m_1+1}(2n)\end{pmatrix}\\\left(1-\frac{B}{2m_2+1}(2n)\right)\end{pmatrix}|2n\rangle. \qquad (7)$$

Here, as above, the superscript (11) indicates two demultiplexed photons and the subscripts $k_1$ and $k_2$ are responsible for the number of photons directed into the measuring modes. In two cases $k_1 = 2m_1, k_2 = 2m_2$ and $k_1 = 2m_1 + 1, k_2 = 2m_2 + 1$ the CV states in equations (4,6) are even while in the two remaining cases $k_1 = 2m_1, k_2 = 2m_2 + 1$ and $k_1 = 2m_1 + 1, k_2 = 2m_2$ the CV states in equations (5,7) are odd.

Their normalization factors become

$$G_{k_1 k_2}^{(11)}(y_2, B) = A_{k_1 k_2\, 0}^{(11)} Z^{(k_1+k_2-2)}(y_2) + \sum_{l=1}^{4} A_{k_1 k_2\, l}^{(11)} \left(y_2 \frac{d}{dy_2}\right)^{l-1} \left(y_2 Z^{(k_1+k_2-1)}(y_2)\right), \quad (8)$$

where the coefficients $A_{k_1 k_2\, l}^{(11)}$ are formed from the internal amplitudes of the measurement induced CV states of a certain parity

$$a_{k_1 k_2\, 0}^{(11)} = 1 - \frac{k_2-1}{k_1} B, \quad a_{k_1 k_2\, 1}^{(11)} = \left(\frac{(k_2-1)B - k_1 - k_2}{k_1 k_2}\right) B, \quad a_{k_1 k_2\, 2}^{(11)} = \frac{B^2}{k_1 k_2} \quad (9)$$

as

$$A_{k_1 k_2\, 0}^{(11)} = a_{k_1 k_2\, 0}^{(11)2}, \quad A_{k_1 k_2\, 1}^{(11)} = 2 a_{k_1 k_2\, 0}^{(11)} a_{k_1 k_2\, 1}^{(11)}, \quad A_{k_1 k_2\, 2}^{(11)} = a_{k_1 k_2\, 1}^{(11)2} + 2 a_{k_1 k_2\, 0}^{(11)} a_{k_1 k_2\, 2}^{(11)},$$
$$A_{k_1 k_2\, 3}^{(11)} = 2 a_{k_1 k_2\, 1}^{(11)} a_{k_1 k_2\, 2}^{(11)}, \quad A_{k_1 k_2\, 4}^{(11)} = a_{k_1 k_2\, 2}^{(11)2}. \quad (10)$$

Differential polynomial in Eq. (8) is formed using $k_1 + k_2 - 1$ derivatives of the analytical function $Z(y_2) = 1/\sqrt{1 - 4y_2^2}$, which depends on $y_2$.

The measurement of $k_1$ and $k_2$ photons in the measuring channels in Figure 1 leads to the generation of measurement induced CV states of a certain parity determined by the expressions (4-7). Knowing the output amplitudes in equation (3), we can estimate the probability distribution of the measurement outcomes

$$P_{k_1 k_2}^{(11)} = \frac{1}{\cosh s} c_{k_1}^{(1)2}(y_1, B) c_{k_2}^{(1)2}(y_2, B) G_{k_1 k_2}^{(11)}(y_2, B), \quad (11)$$

thereby obtaining the probabilities of the measurement induced generation of the corresponding CV states of a certain parity.

We have considered measurement induced quantum engineering of new nonclassical CV states of light. Here we are going to adapt the nonclassical states to quantum engineering even/odd SCSs which are superpositions of coherent states with equal in magnitude but opposite in sign amplitudes, that is, $\beta$ and $-\beta$

$$|SCS_+(\beta)\rangle = N_+(|\beta\rangle + |-\beta\rangle) = \sum_{n=0}^{\infty} b_{2n}^{(+)} |2n\rangle, \quad (12)$$

$$|SCS_-(\beta)\rangle = N_-(|\beta\rangle - |-\beta\rangle) = \sum_{n=0}^{\infty} b_{2n+1}^{(-)} |2n+1\rangle, \quad (13)$$

where their amplitudes in the Fock representation are given by $b_{2n}^{(+)} = 2 N_+ \exp(-\beta^2/2) \left(\beta^{2n}/\sqrt{(2n)!}\right)$ and $b_{2n+1}^{(-)} = 2 N_- \exp(-\beta^2/2) \beta^{2n+1}/\sqrt{(2n+1)!}$ with $N_\pm = \left(2(1 \pm \exp(-2\beta^2))\right)^{-1/2}$.

A measure of the closeness between the target SCSs and the measurement induced CV states can be the fidelity, which in the Fock representation is defined as $Fid_{2m_1\, 2m_2}^{(11)} = \left(\sum_{n=0}^{\infty} b_{2n}^{(+)} b_{2m_1\, 2m_2}^{(11)}\right)^2$, $Fid_{2m_1+1\, 2m_2+1}^{(11)} = \left(\sum_{n=0}^{\infty} b_{2n}^{(+)} b_{2m_1+1\, 2m_2+1}^{(11)}\right)^2$ for even SCS and $Fid_{2m_1+1\, 2m_2}^{(11)} = \left(\sum_{n=0}^{\infty} b_{2n}^{(+)} b_{2m_1+1\, 2m_2}^{(11)}\right)^2$, $Fid_{2m_1\, 2m_2+1}^{(11)} = \left(\sum_{n=0}^{\infty} b_{2n}^{(+)} b_{2m_1\, 2m_2+1}^{(11)}\right)^2$ for odd SCS, where the amplitudes $b_{2m_1\, 2m_2}^{(11)}$, $b_{2m_1+1\, 2m_2}^{(11)}$, $b_{2m_1\, 2m_2+1}^{(11)}$ and $b_{2m_1+1\, 2m_2+1}^{(11)}$ used follow directly from the definition of the measurement induced CV states of a certain parity in the equations (4-7). Since the amplitudes $b_{k_1 k_2}^{(11)}(B, S)$ depend on two parameters $B$ and $S$, the fidelity $Fid_{k_1 k_2}^{(11)}(B, S, \beta)$ is determined by three parameters $B, S$ and $\beta$ by varying which one can achieve the highest possible fidelity close to 1. Those values of $B$ and $S$ at which the fidelity reaches its maximum value for a given amplitude of $\beta$ we are going to call optimizing and denote them as $B^{(11)}_{k_1k_2\,opt}$ and $S^{(11)}_{k_1k_2\,opt}$.

The number of possible cases of realization of the SCSs is quite large since an arbitrary set of measured photons $k_1$ and $k_2$ can ensure the generation of the target state. For even SCSs we chose the outcomes with the same number of measured photons in both measurement channels, i.e. $k_1 = k_2 = k$. In figure 2(a) we show the fidelity of the measurement induced CV state optimized by parameters $B$ and $S$ as a function of the target state amplitude $\beta$ for $k = 2,4,6$. In general, it is observed that increasing the number of subtracted photons allows generating the target state with greater fidelity. Thus, in the case of $k = 6$, the fidelity$> 0.99$ is observed for the amplitude of the target state at least more than 2.5, i.e. $\beta > 2.5$. The optimizing values $B^{(11)}_{kk\,opt}$ and $S^{(11)}_{kk\,opt}$ that provide the highest possible fidelity of the measurement induced CV state with the target even SCS as a function of $\beta$ are shown in Figures 2(b,c), respectively. As for the optimizing $B^{(11)}_{kk\,opt}$, two beam splitters with a larger transmittance $T > R$ are required to achieve maximum fidelity. But it is worth noting here that the used BSs with $0.05 < B^{(11)}_{kk\,opt} < 0.9$ are not highly transmissive beam splitters, when the BS operator can be approximated by a set of corresponding operators in neighboring modes. It is important to note that increasing the number of subtracted demultiplexed photons allows to reduce the requirement for initial squeezing as shown in Figure 2(c), that is, $S^{(11)}_{22\,opt} > S^{(11)}_{44\,opt} > S^{(11)}_{66\,opt}$ and this reduction can be significant. This finding allows experimenters to redirect their efforts to increasing the number of subtracted photons instead of increasing the initial squeezing. Finally, in Figure 2(d) we present the graphs of the dependence of probabilities $P^{(11)}_{kk}\left(B^{(11)}_{kk\,opt}, S^{(11)}_{kk\,opt}\right)$ on the amplitude $\beta$ of the target state. These probabilities are obtained for those $B^{(11)}_{kk\,opt}$ and $S^{(11)}_{kk\,opt}$ at which the maximum fidelity between the measurement induced and target states is detected in Fig. 2(a). A distinctive feature of the mechanism is the increased probability value to $> 0.04$ for $P^{(11)}_{22}$, which for the initial superposition, in which the vacuum state prevails, is undoubtedly an important consequence. The increase in the probability of measurement outcomes occurs due to the pumping the initial system in Fig. 1 by two demultiplexed photons through a system of beam splitters.

As for quantum engineering of odd SCS, we add one photon to the second measuring channel to the measurement outcomes used for the construction in figure 2, that is, we consider the following measurement results $k_1 = 2, k_2 = 3, k_1 = 4, k_2 = 5$ and $k_1 = 6, k_2 = 7$. The corresponding dependencies of the maximum possible fidelity $Fid^{(11)}_{k_1k_2}\left(B^{(11)}_{k_1k_2\,opt}, S^{(11)}_{k_1k_2k\,opt}\right)$ of the measurement induced CV state approximating the odd SCS, values of $B^{(11)}_{k_1k_2\,opt}$ and $S^{(11)}_{k_1k_2k\,opt}$, which optimize the fidelity as well as the probabilities $P^{(11)}_{k_1k_2}\left(B^{(11)}_{k_1k_2\,opt}, S^{(11)}_{k_1k_2k\,opt}\right)$ of their generation are presented in Figure 3(a-d), respectively. In general, these dependencies have some similarity with those shown in Figure 2(a-d). The difference is only in quantitative values. So, the fidelity of odd SCS is taken to be slightly higher than those shown in figure 2(a). But the probability of measurement induced generation of the odd CV states may be slightly less than in the previous case.

It is interesting to compare the results presented in Figures 2 and 3 with those obtained in the case of vacuum inputs in beam splitters [29] as if we used $|00\rangle_{23}$ at the input to the beam splitters instead of $|11\rangle_{23}$ in figure 1. The corresponding dependences of the fidelity both with $Fid^{(11)}_{k_1k_2}$ and without $Fid^{(00)}_{k_1k_2}$ demultiplexed photons with the same number of subtracted photons on the amplitude $\beta$ of the target even/odd SCSs are shown in figures 4(a,b) for even and odd CV states, respectively. As can be seen from the presented dependencies, $Fid_{k_1k_2}^{(11)} > Fid_{k_1k_2}^{(00)}$ are found, which indicates in favor of using input demultiplexed photons to increase the fidelity of the output CV state. The only exception is the case with $k_1 = k_2 = 2$, where the fidelity $Fid_{22}^{(00)}$ is not much better than the fidelity $Fid_{22}^{(11)}$, i.e. $Fid_{22}^{(00)} > Fid_{22}^{(11)}$. In general, it is found that the greater the number of subtracted photons, the more the difference in the fidelity of the output CV states. To quantitatively evaluate the gain of the fidelity in CV output state, we consider the following the following parameter $g_{k_1k_2}^{(00,11)} = 10log_{10} Fid_{k_1k_2}^{(11)}/Fid_{k_1k_2}^{(00)}$ expressed in $dB$ the dependencies of which on $\beta$ are presented in figure 4(c,d). Parameter $g_{k_1k_2}^{(00,11)}$ acquires a small but positive value except $g_{22}^{(00,11)}$ in agreement with the results presented in the figure 4(a,b). Increasing the number of subtracted photons leads to gain of the fidelity of the output state.

As for the possibility of increasing the probability of quantum engineering SCSs by using two input demultiplexed photons, the corresponding dependencies of the gain $j_{k_1k_2}^{(00,11)} = 10log_{10} P_{k_1k_2}^{(11)}/P_{k_1k_2}^{(00)}$ on $\beta$ are presented in Figure 5(a,b) for even and odd cases. It is worth noting here that following [29] the fidelity $Fid_{k_1k_2}^{(00)}$ depends only on one parameter $y_2$, and not on two, as in the case of two additional input photons. This circumstance directly relates the initial squeezing $y_2$ and the BS parameter $B$ as $y = y_2(1 + B)^2$, where $y_2$ follows from the fidelity optimization of $Fid_{k_1k_2}^{(00)}(y_2)$. This allows to increase the input squeezing $y > y_2$ by increasing $B$, which directly affects the probability of being proportional to the product $y_2B$. In other words, the case without additional input photons allows increasing the probability $P_{k_1k_2}^{(00)}$ of a positive outcome by increasing the initial squeezing $S$ and $B$ while maintaining the maximum possible fidelity $Fid_{k_1k_2\ max}^{(00)}$, which is used to construct the dependencies in Figure 5(a,b). In Figure 5(a,b), the initial $S$ is chosen to be very large, significantly exceeding the practically achievable level of squeezing. Even despite the choice of very large values of $S \gg 10\ dB$ for the calculation of the gain $j_{k_1k_2}^{(00,11)}$ in the practically important case of $\beta > 1$ the gain takes on values calculated in tens. The choice of such a large $S \gg 10\ dB$ is reflected in the presence of negative values $j_{k_1k_2}^{(00,11)} < 0$ for practically inessential values $\beta < 1$. Reducing the initial squeezing to acceptable values $S < 10\ dB$ allows the function $j_{k_1k_2}^{(00,11)}$ to be realized positively in the entire range of change $\beta$. Then the peak values of the function $j_{k_1k_2}^{(00,11)}$ are counted in hundreds. Comparing the increase in the fidelity and probability when adding two demultiplexed photons, it becomes clear that the gain in probability is more significant.

## 3. Conclusion

We have proposed and studied in detail the influence of input demultiplexed photons on the efficiency of quantum engineering of even/odd SCSs of large amplitude $\beta > 2.5$ using the SMSV state as the initial ingredient of the protocol. In the consideration, the model of a realistic beam splitter is used [8,29-31], which allows optimizing the output characteristics of light by varying not only the initial squeezing but also the BS parameter. This model provides a significant advantage over the simplified version that is based on replacing the beam splitter operator with a set of creation and annihilation operators, an approximation that is used in the case of high transmittance beam splitter [7]. Indeed, an increase of the reflection coefficient of the beam splitter enables at least to magnify the probability of measurement outcomes and thus the generation of the measurement induced CV states of light compared to the case of highly transmissive beam splitter [8].

The combined action of adding two input photons and subtracting a certain number of photons allows to increase the fidelity$> 0.99$ of the output measurement induced CV state of a certain parity. The observed fidelity of the CV states of the with two input photons exceeds the maximum possible fidelity in the absence of photons. Numerical results show that the more photons are subtracted, the greater the fidelity of the output state. Moreover, subtracting more photons mitigates the requirements for the initial squeezing of the SMSV state. Demultiplexing, i.e. directing two input photons into two different modes and then measuring the number of photons in two measurement channels, can be an advantage over having the input and output photons in the same mode [41]. Adding two initial photons also increases the probability of generating the required measurement induced CV states, both due to the total number of photons involved in the mixing and due to the increase in the BS reflectivity. This increase in probability is very strong more than $100\ dB$ even if the original SMSV state with very high squeezing is used. Using SMSV state with practically realizable squeezing further enhances the gain of the probability. All this fully justifies the addition of two input demultiplexed photons.

The proposed scheme with two additional photons is feasible in practice. The number of photons subtracted is within the capabilities of existing photon number resolving technology. [41]. Special cases of coinciding measurements with the same number of photons are considered, although the scheme allows for quantum engineering of even/odd SCSs with $k_1 \ne k_2$. The scheme is scalable and allows for various types of modification, which is the subject of new separate studies.

**Acknowledgement**

The study was supported by the grant of the Russian Science Foundation No. 25-12-20026, https://rscf.ru/project/25-12-20026/.

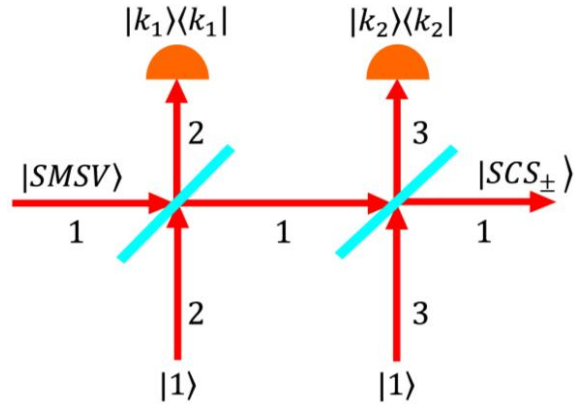

**Figure 1** Schematic representation of scheme of the quantum engineering of even/odd SCSs states by subtracting photons from the initial SCS state (mode 1) realized by two PNR detectors. Two input demultiplexed photons in auxiliary modes (modes 2 and 3) are used to increase the efficiency of quantum engineering, which is expressed in gain of the fidelity of the output state, as well as in an increase in the probability of the realization of such measurement induced CV states.

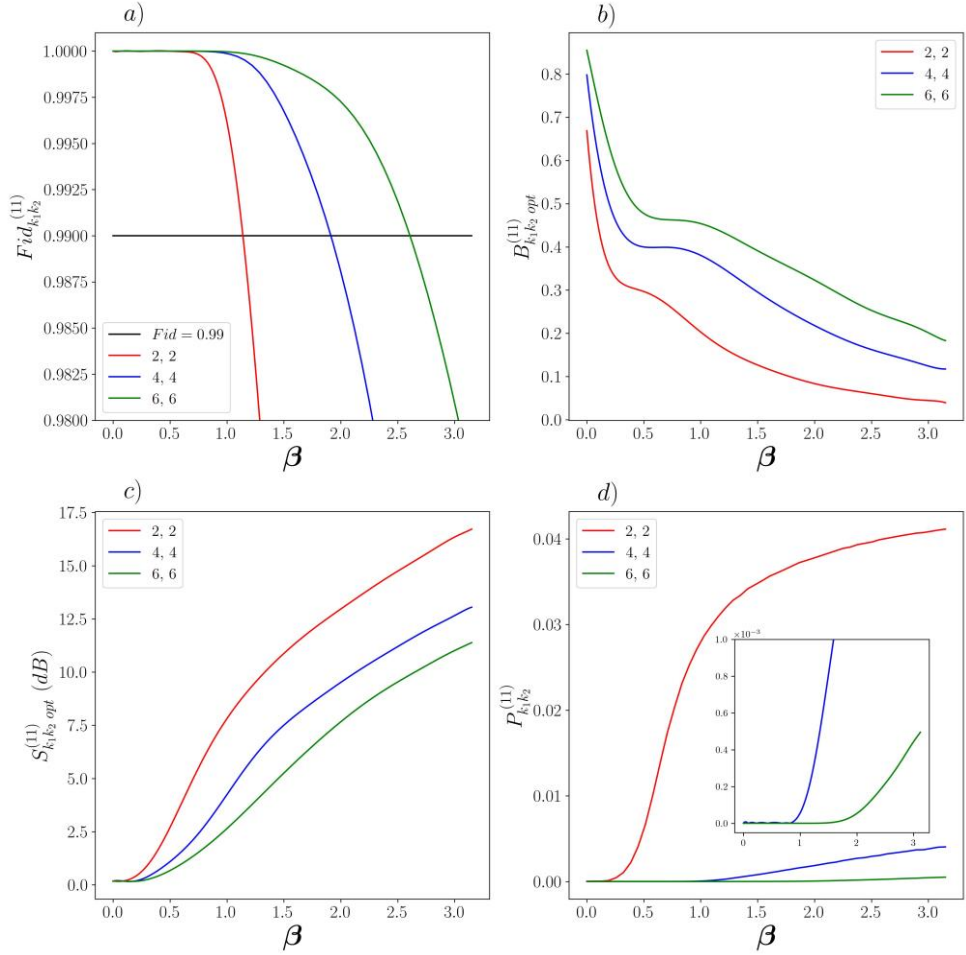

**Figure 2(a-d)** (a) Dependence of the optimized fidelity $Fid^{(11)}_{k_1 k_2}\left(B^{(11)}_{k_1 k_2\ opt}, S^{(11)}_{k_1 k_2\ opt}\right)$ of the measurement induced CV state of a certain parity with respect to the target even SCS on its amplitude $\beta$ for coinciding numbers of photons $k_1 = k_2 = 2$, $k_1 = k_2 = 4$ and $k_1 = k_2 = 6$ measured in both channels. The horizontal line corresponds to the fidelity level of 0.99. The dependences of the optimizing parameters $B^{(11)}_{k_1 k_2\ opt}$ and $S^{(11)}_{k_1 k_2\ opt}$ on the SCS amplitude $\beta$ are presented in (b) and (c), respectively. The same optimizing parameters determine the probability $P^{(11)}_{k_1 k_2}\left(B^{(11)}_{k_1 k_2\ opt}, S^{(11)}_{k_1 k_2\ opt}\right)$ of measurement induced generation of the CV states as function of $\beta$ in (d). The inset shows the dependences $P^{(11)}_{44}$ and $P^{(11)}_{66}$ since in the scale used they look like almost horizontal lines.

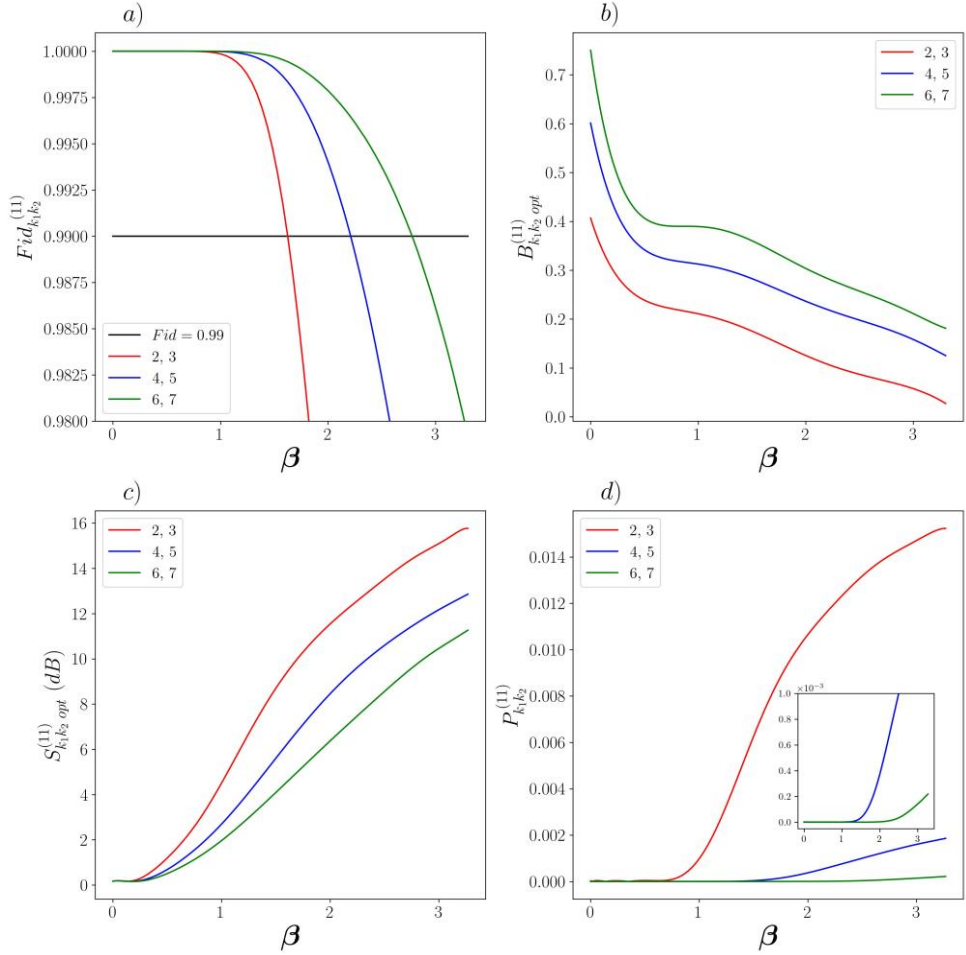

**Figure 3(a-d)** Dependences (a) $Fid^{(11)}_{k_1k_2}\left(B^{(11)}_{k_1k_2\ opt}, S^{(11)}_{k_1k_2\ opt}\right)$, (b) $B^{(11)}_{k_1k_2\ opt}$, (c) $S^{(11)}_{k_1k_2\ opt}$ and $P^{(11)}_{k_1k_2}\left(B^{(11)}_{k_1k_2\ opt}, S^{(11)}_{k_1k_2\ opt}\right)$ on $\beta$ as in Figure 2 but for odd CV states. The total number of subtracted photons $k_1 = 2, k_2 = 3$, $k_1 = 4, k_2 = 5$ and $k_1 = 6, k_2 = 7$ is odd, which guarantees the generation of odd CV states. The insert in (d) is made on a smaller scale to show the details of $P^{(11)}_{45}$ and $P^{(11)}_{67}$.

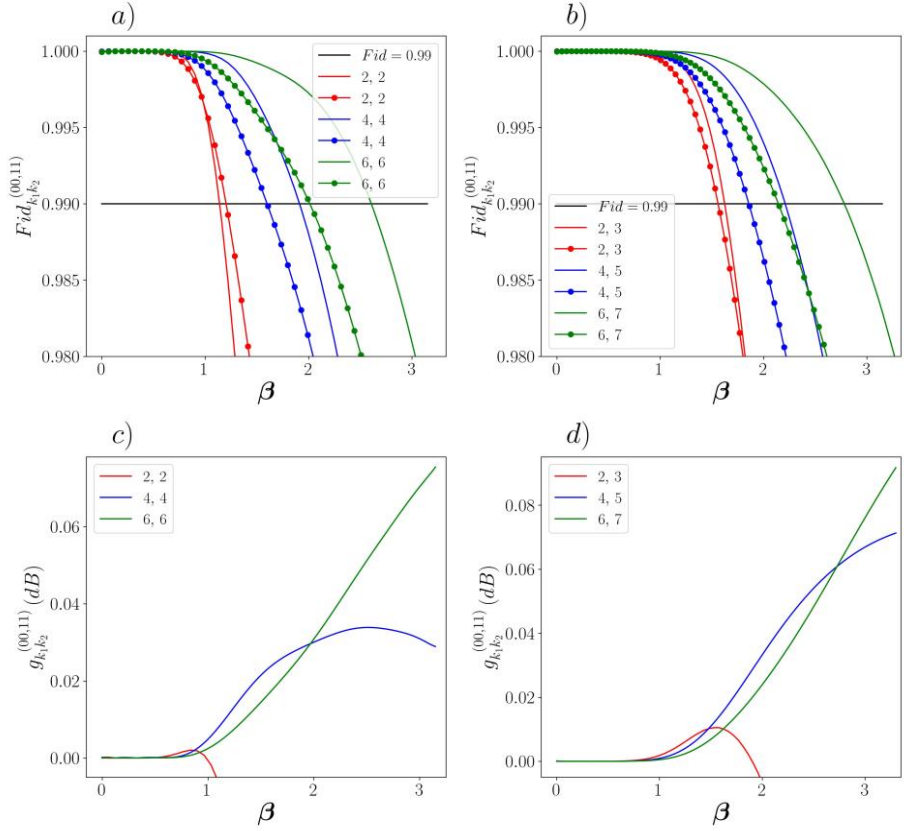

**Figure 4(a-d)** Comparison of the fidelities between output measurement induced CV states and (a) even and (b) odd SCSs in the case of two demultiplexed photons (solid line) and without their use (solid lines with additional dots). It is obvious that the output fidelity with two input photons $Fid^{(11)}_{k_1 k_2}$ is superior to the fidelity $Fid^{(00)}_{k_1 k_2}$ of the measurement induced CV state without using additional photons except $Fid^{(00)}_{22} > Fid^{(11)}_{22}$. The gain in the fidelity $g^{(00,11)}_{k_1 k_2} = 10 log_{10} Fid^{(11)}_{k_1 k_2} / Fid^{(00)}_{k_1 k_2}$ expressed in $dB$ on amplitude $\beta$ when using two demultiplexed photons is presented in (c) for even and in (d) for odd SCSs.

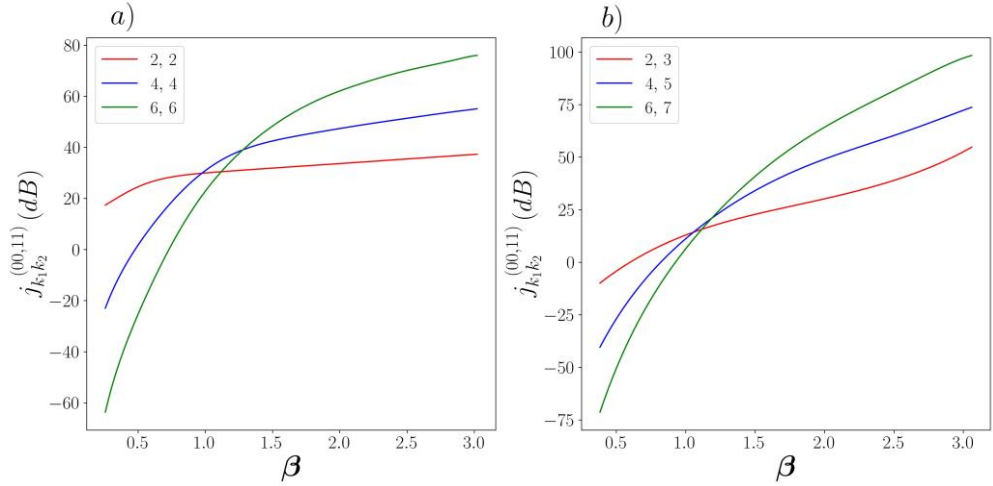

**Figure 5(a,b)** Comparison of the probabilities of generating target (a) even and (b) odd SCSs both in the absence and in the presence of two demultiplexed photons, presented through the evolution of the function $j_{k_1 k_2}^{(00,11)}$ from $\beta$ in the case of a practically unachievable initial squeezing $S \gg 10 \, dB$. Due to the choice of a very large initial squeezing, negative values of $j_{k_1 k_2}^{(00,11)} < 0$ occur in a practically insignificant region $\beta < 1$. However, in the more significant region $\beta > 1$ the gain of the probability acquires quite large values even up to $100 \, dB$. When the initial squeezing decreases to practically realizable values $S < 10 \, dB$, the gain becomes positive for any values $\beta$ with an increase in peak values.